\documentclass[pra,onecolumn, showpacs, showkeys, secnumarabic, aps, amsmath, amssymb, nofootinbib, superscriptaddress, longbibliography, floatfix, table-of-contents, dblfloatfix]{revtex4-2}
\usepackage[utf8]{inputenc}
\usepackage[pdftex]{graphicx}
\usepackage{mathrsfs}
\usepackage[colorlinks, breaklinks, urlcolor={blue}, linkcolor={blue}, citecolor={blue}]{hyperref}
\usepackage{array}
\usepackage{amsmath}
\usepackage{type1cm}
\usepackage[english]{babel}
\usepackage{lmodern}
\usepackage{microtype}
\usepackage{booktabs}
\usepackage{caption}
\usepackage{braket}
\usepackage{xcolor}
\usepackage{orcidlink}
\usepackage{tikz}

\begin{document}
\title{Basis-Independent Coherence Dynamics of Tripartite States under Pure Dephasing}

\author{Sovik Roy \orcidlink{0000-0003-4334-341X} }
\email[]{s.roy2.tmsl@ticollege.org}
\affiliation{Department of Mathematics, Techno Main Salt Lake (Engg. Colg.), \\Techno India Group, EM 4/1, Sector V, Salt Lake, Kolkata  700091, India}

\author{Abhijit Mandal\orcidlink{0000-0001-7101-9495}}
\email[]{a.mandal1.tmsl@ticollege.org}
\affiliation{Department of Mathematics, Techno Main Salt Lake (Engg. Colg.), \\Techno India Group, EM 4/1, Sector V, Salt Lake, Kolkata  700091, India}

\begin{abstract}
\noindent Quantum coherence is a fundamental quantum resource whose preservation under environmental interactions is essential for quantum information processing. While most studies have focused on basis-dependent coherence measures, the dynamics of intrinsic coherence quantified by basis-independent measures remain largely unexplored. In this work, we investigate the dynamics of basis-independent quantum coherence for several representative tripartite pure and mixed states subjected to local and common dephasing environments in both Markovian and non-Markovian regimes. We show that Markovian local dephasing leads to state- dependent coherence degradation, whereas collective dephasing significantly enhances coherence preservation through decoherence-free sectors. More importantly, non-Markovian environments give rise to nearly frozen coherence dynamics for both pure and mixed states, demonstrating the remarkable robustness of intrinsic coherence against dephasing environment. A comparison with the measure of relative entropy of coherence reveals that basis-independent coherence measure exhibits substantially greater resilience and qualitatively different dynamical behaviour than its basis-dependent counterpart. These results provide new insights into the preservation of intrinsic multipartite coherence in open quantum systems and highlight basis-independent coherence as a robust quantum resource for realistic noisy quantum technologies.
\end{abstract}


\maketitle

\section{Introduction}
\label{sec:intro}
\noindent One of the most fundamental manifestations of non-classicality is the quantum coherence originates from the superposition principle. It is one of the indispensable resources in quantum information science \cite{Baumgratz2014,Streltsov2017,Chitambar2019}. The presence of coherence enables several quantum advantages in diverse fields such as quantum computation and communication \cite{fs2013}, quantum metrology \cite{dpp2018}, quantum thermodynamics \cite{pk2016}, quantum cryptography \cite{bhut1995} and many more \cite{ramirez2023,mh2016,heinrich21}. Consequently, the quantification and characterization of quantum coherence have attracted considerable attention in recent years and have led to the developments of rigorous resource theoretic framework of coherence. Several measures for quantifying quantum coherence have been proposed. Among them relative entropy of coherence and $l_1$-norm of coherence are the most extensively used, owing to their simple mathematical structure as well as clear operational interpretation \cite{hz2018,srana2017,zhu2018}. Despite their widespread adoption, these coherence measures are inherently basis dependent. A prescribed reference basis is, what is needed for applying these measures. While such basis dependence can be advantageous when a physically distinguished basis naturally exists, e.g. computational basis or the energy eigen-basis, it obscures the interpretation of coherence as an intrinsic property of a given quantum state. Therefore, a basis-independent characterization of coherence is highly desirable for understanding the fundamental coherence properties of multipartite quantum systems. To overcome this limitation, a basis-independent measure of quantum coherence based on the quantum Jensen-Shannon divergence between a quantum state and the maximally mixed state has recently been introduced \cite{Radhakrishnan2019}. The resulting measure is invariant under arbitrary unitary transformations and therefore quantifies the intrinsic coherence of a quantum state, independent of the choice of the basis. Depending on its metric properties and operational significance, the basis-independent measure proposed by Radhakrishnan \textit{et al.} \cite{Radhakrishnan2019} provides a natural framework for comparing the coherence properties of quantum states possessing entirely different superposition structures.\\ 

\noindent Now, the decoherence is the process by which the state of a quantum system loses its ability to exhibit coherent superposition due to interactions with its surrounding environment. The unavoidable interaction of the state of a realistic quantum system with its surrounding environment leads to decoherence and consequently degrades quantum resources \cite{Breuer2002,Rivas2012,Zurek2003}. Therefore, the theory of open quantum systems plays a central role in understanding the dynamical behaviour of quantum coherence under environmental dynamics. Of particular importance are dephasing environments, which suppress phase correlations while leaving the populations of the quantum states unchanged. In quantum mechanics, the population of a state is simply the probability that the system occupies that state. Among the various decoherence mechanisms, pure dephasing is particularly well suited for the present investigation because it selectively suppresses quantum coherence without inducing energy relaxation. As the populations remain unchanged throughout the evolution, the dynamics are governed solely by the decay of phase correlations, enabling a clear assessment of the robustness of the coherence measure itself. This is especially important when comparing basis-dependent and basis-independent coherence measures, since the observed differences can be attributed primarily to the properties of the quantifiers rather than to simultaneous changes in both populations and coherences that occur in dissipative channels such as amplitude damping or depolarizing noise. Nevertheless, depending on the nature of the \textit{system-environment interaction}, the dynamics may be Markovian or non-Markovian \cite{merkil2022,hprev2016}. In Markovian processes, information flows irreversibly from the system to the environment and coherence decays monotonically. In contrast, non-Markovian environments possess memory effects and may induce partial restoration and revival of quantum coherence through the backflow of information from the environment to the system \cite{Breuer2009,Rivas2014}. Moreover, environmental interactions may occur independently through local reservoirs or collectively through common reservoirs, leading to substantially different coherence dynamics and, in certain situations, the emergence of decoherence-free subspaces \cite{Lidar1998,Zanardi1997}. In a recent study, one of the present authors investigated the dynamics of tripartite coherence based on the measure of relative entropy of coherence under dephasing environments \cite{roy2025}. Motivated by these findings, we investigate the dynamics of tripartite coherence using the basis-independent coherence measure proposed by Radhakrishnan \textit{et al.} \cite{Radhakrishnan2019}. This enables a direct comparison with the corresponding basis-dependent dynamics and demonstrates that intrinsic tripartite coherence is remarkably resilient to dephasing, exhibiting nearly frozen behaviour under non-Markovian evolution.\\

\noindent  Tripartite pure entangled states provide an ideal test-bed for investigating the interplay between intrinsic coherence and environmental interactions. Among the most prominent examples are the Greenberger-Horne-Zeilinger ($|GHZ\rangle$) \cite{GHZ1989} and $|W\rangle$ \cite{Dur2000} states, which constitute inequivalent classes of genuine tripartite entanglement. In addition to that, the $|W\overline{W}\rangle$ state and the $|Star\rangle$ state \cite{cao2020} exhibit distinct superposition structures and robustness properties under decoherence. Moreover, the mixed states constructed by taking convex combinations of pure tripartite states ($|GHZ\rangle$ and $|W\rangle$), convex combination of pure tripartite state and maximally mixed state or white noise (viz. $|Werner\rangle - |GHZ\rangle$ and $|Werner\rangle - |W\rangle$), further enrich the investigation by providing a continuous interpolation between highly coherent quantum states and completely mixed states. These states have earlier been studied for studying various dynamical nature in open quantum system \cite{roy2024,roy2025(1)}.\\

\noindent In particular, it remains unclear how different classes of three-qubit pure and mixed states preserve their intrinsic coherence under local and common dephasing environments in both Markovian and non-Markovian regimes. Motivated by these considerations, in the present work we investigate the dynamics of basis-independent quantum coherence for the above important classes of states. Employing the basis-independent coherence measure, we examine the evolution of intrinsic coherence under local and collective dephasing channels in both Markovian and non-Markovian environments. Particular attention is devoted to understanding the robustness hierarchy of these class of states, the role of environmental memory effects, and the occurrence of decoherence-free behaviour under collective interactions. The present study therefore provides a comprehensive characterization of intrinsic tripartite coherence and highlights the fundamental distinctions between basis-dependent and basis-independent descriptions of quantum coherence in open quantum systems. To the best of our knowledge, this is the first systematic investigation of basis-independent coherence dynamics for representative tripartite pure and mixed states under both local and common Markovian and non-Markovian dephasing environments.\\

\noindent The paper is organized as follows. In \textcolor{blue}{Sec.~$II$}, we describe the theoretical framework of the open quantum system and introduce the Markovian and non-Markovian local and common dephasing models for a system of three non-interacting qubits. In \textcolor{blue}{Sec.~$III$}, we briefly review the basis-independent measure of quantum coherence based on the quantum Jensen-Shannon divergence and discuss its fundamental properties. \textcolor{blue}{Sec.~$IV$} is devoted to the dynamics of basis-independent coherence for various classes of tripartite pure states, namely the $|GHZ\rangle$, $|W\rangle$, $|W\overline{W}\rangle$, and $|Star\rangle$ states, under different dephasing environments. In \textcolor{blue}{Sec.~$V$}, we investigate the coherence dynamics of mixed tripartite states, including the $|GHZ\rangle - |W\rangle$ mixture, $|Werner\rangle - |GHZ\rangle$ state, and $|Werner\rangle - |W\rangle$ state. textcolor{blue}{Sec.~$VI$}, presents the effect of partial exposure to the dephasing environment, in which one qubit is isolated from local Markovian noise. Finally, the main findings of the present work are summarized in \textcolor{blue}{Sec.~$VII$}.

\section{Theoretical Framework}

\noindent The unavoidable interaction between a quantum system and its surrounding environment leads to decoherence and consequently degrades quantum resources such as entanglement and coherence. In realistic situations, the dynamics of a multipartite quantum system is therefore more appropriately described within the framework of open quantum systems. Among various environmental interactions, pure dephasing channels occupy a central position since they destroy phase correlations while preserving the populations of the energy eigenstates. The resulting loss of phase information directly influences the coherence properties of multipartite quantum states.\\

\begin{figure}
    \centering
    \includegraphics[width=0.8\linewidth]{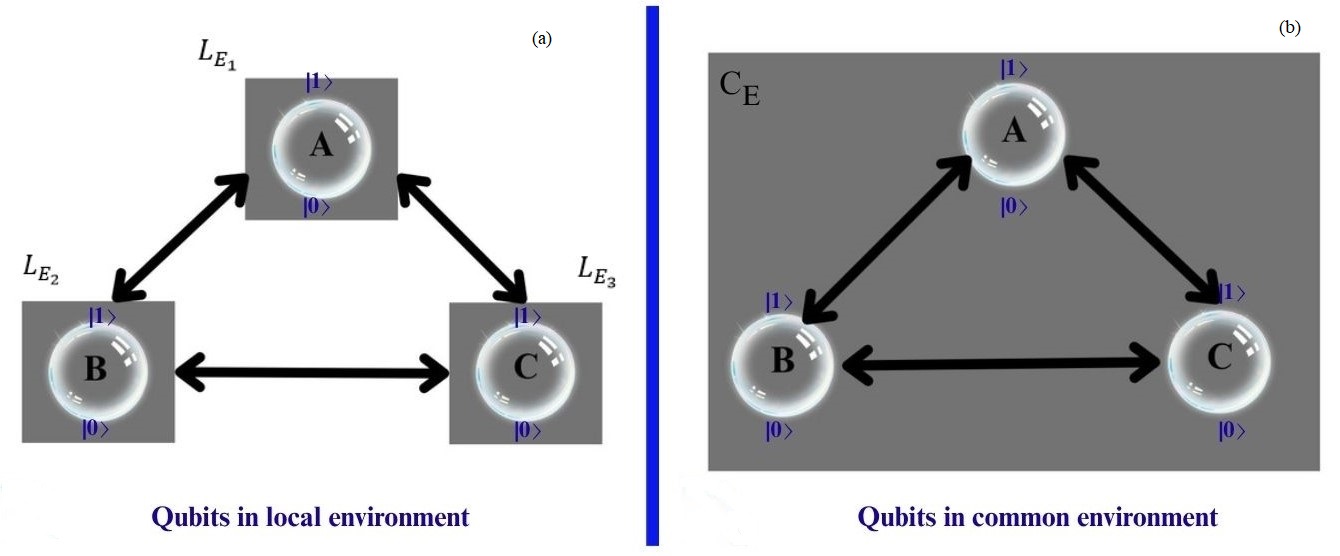}
    \caption{Schematic representation of the two dephasing models considered in the present work. (a) Local dephasing model: the three non-interacting qubits $A$, $B$, and $C$ are independently coupled to their respective bosonic reservoirs $L_{E_1}$, $L_{E_2}$, and $L_{E_3}$. The environmental fluctuations are therefore uncorrelated. (b) Common dephasing model: three qubits are collectively coupled to a single bosonic reservoir $C_E$, giving rise to correlated environmental fluctuations and possible decoherence-free behaviour. These models form the basis for investigating the dynamics of basis-independent quantum coherence under Markovian and non-Markovian environments.}
    \label{fig:placeholder}
\end{figure}

\noindent In the present work, we consider a system consisting of three non-interacting qubits subjected to two physically distinct dephasing scenarios. In the first scenario (\textcolor{blue}{depicted in Fig.~$1(a)$}), each qubit interacts independently with its own bosonic reservoir (qubit $A$ in $L_{E_1}$, qubit $B$ in $L_{E_2}$ and qubit $C$ in $L_{E_3}$),leading to local dephasing. In the second scenario (\textcolor{blue}{shown in Fig.~$1(b)$}), three qubits are coupled collectively to a single bosonic reservoir ($C_E$), resulting in common dephasing. Depending upon the environmental correlation time and the system-reservoir interaction strength, both models can exhibit either Markovian or non-Markovian dynamics.\\

\noindent The consideration of both local and common dephasing environments is essential because they represent physically distinct system--environment configurations. In the local model, each qubit interacts with an independent reservoir and therefore experiences statistically independent phase fluctuations. In the common model, all qubits interact collectively with a single reservoir and consequently experience correlated environmental fluctuations. These correlations can generate decoherence-free sectors and substantially modify the robustness of multipartite coherence. By comparing the two models, we can therefore distinguish the effects of independent environmental decoherence from those arising from correlated environmental interactions. This comparison is particularly relevant for multipartite states, for which the structure of the superposition determines their sensitivity to collective noise.

\subsection{three-qubits Under Local Dephasing}

\noindent We first consider a system of three non-interacting qubits, labelled by $A$, $B$, and $C$, each of which is independently coupled to its own bosonic environment. Such a configuration represents a local dephasing scenario and is schematically illustrated in \textcolor{blue}{Fig.~$1(a)$}. Since the three reservoirs do not communicate with one another, the environmental fluctuations acting on different qubits are statistically independent. The total Hamiltonian describing the system and the three local reservoirs can be written as \cite{Breuer2002,Rivas2012}

\begin{equation}
H=
\sum_{i=1}^{3}
\left(
\frac{\omega_0}{2}\sigma_z^{(i)}
+
\sum_k \omega_{ik} b_{ik}^{\dagger}b_{ik}
+
\sigma_z^{(i)}
\left(
B_i+B_i^{\dagger}
\right)
\right),
\end{equation}

\noindent where $\sigma_z^{(i)}$ denotes the Pauli spin operator of the $i^{th}$ qubit and $\omega_0$ is the transition frequency of each qubit. The operators $b_{ik}$ and $b_{ik}^{\dagger}$ are respectively the annihilation and creation operators corresponding to the $k$th mode of the reservoir coupled to the $i^{th}$ qubit. The environmental operator is defined by

\begin{equation}
B_i=\sum_k g_{ik}b_{ik},
\end{equation}

\noindent where $g_{ik}$ denotes the coupling strength between the $i^{th}$ qubit and the $k^{th}$ mode of its corresponding reservoir. The Hamiltonian consists of three physically distinct contributions. The first term represents the free Hamiltonian of the three-qubit system, the second term corresponds to the Hamiltonian of the bosonic reservoirs, and the third term describes the system-environment interaction responsible for dephasing. Since the interaction Hamiltonian is proportional to
$\sigma_z^{(i)}$, it commutes with the system Hamiltonian,

\[
[H_S,H_{SB}]=0.
\]

\noindent Consequently, no exchange of energy takes place between the qubits and their environments. The populations of the computational basis states remain invariant during the evolution and only the phase relations, represented by the off-diagonal elements of the density matrix, decay with time. The resulting dynamics therefore corresponds to a pure dephasing process.
Assuming that the total system-reservoir state is initially factorized,

\[
\rho_{\rm tot}(0)
=
\rho(0)\otimes
\rho_{E_1}\otimes
\rho_{E_2}\otimes
\rho_{E_3},
\]

\noindent and that each reservoir is initially prepared in a thermal equilibrium state at temperature $T_i$, the reduced density operator of the three-qubit system satisfies the master equation

\begin{equation}
\frac{d\rho(t)}{dt}
=
-i[H_S,\rho(t)]
+
\sum_{i=1}^{3}
\gamma_i(t)
\left(
\sigma_z^{(i)}
\rho(t)
\sigma_z^{(i)}
-
\rho(t)
\right),
\end{equation}

\noindent where

\begin{equation}
H_S
=
\frac{\omega_0}{2}
\sum_{i=1}^{3}
\sigma_z^{(i)}
\end{equation}

\noindent is the free Hamiltonian of the isolated three-qubit system.\\

\noindent The quantities $\gamma_i(t)$ are the time-dependent dephasing rates and are completely determined by the spectral properties of the reservoirs,

\begin{equation}
\gamma_i(t)
=
2
\int_0^{\infty}
d\omega\,
J_i(\omega)
\coth
\left(
\frac{\omega}{2k_B T_i}
\right)
\frac{\sin(\omega t)}{\omega},
\end{equation}

\noindent where $J_i(\omega)$ denotes the spectral density of the reservoir coupled to the $i^{th}$ qubit. Here, $\omega$ denotes the frequency of an individual reservoir mode, while $\eta_i$ is the cut-off frequency that characterizes the bandwidth of the reservoir.\\

\noindent Throughout the present work, we assume that each local reservoir possesses an Ohmic spectral density,

\begin{equation}
J_i(\omega)
=
\eta_i
\omega
\exp
\left(
-\frac{\omega}{\Lambda_i}
\right),
\end{equation}

\noindent where $\eta_i$ and $\Lambda_i$ respectively denote the system-environment coupling strength and the cut-off frequency. The latter determines the reservoir correlation time ($\tau_B\sim \frac{1}{\Lambda_i}$) and therefore governs the transition between Markovian and non-Markovian dynamics.\\

\noindent The cut-off frequency determines the correlation time of the reservoir. When the environmental correlation time is negligibly small compared to the characteristic time scale of the system, the reservoirs possess no memory and the dephasing rates become effectively constant. The dynamics then reduces to the Markovian regime and coherence decays irreversibly due to a one-way flow of information from the system to the environments.\\

\noindent Conversely, finite values of the cut-off frequency endow the reservoirs with memory and the dephasing rates become explicitly time dependent. The resulting evolution is non-Markovian and is characterized by a temporary backflow of information from the environments to the system, which may significantly suppress coherence degradation and even lead to partial recovery of quantum coherence.

\subsection{Three qubits Under Common Dephasing}

\noindent We next consider a collective dephasing model in which three qubits interact simultaneously with a single bosonic reservoir, as shown in Fig.~$1(b)$. Unlike the local dephasing model, all qubits are now exposed to the same environmental degrees of freedom. Consequently, the fluctuations experienced by different qubits become correlated and may induce collective effects that are absent in independent environments. Unlike independent reservoirs, a common environment may preserve specific collective superposition states, giving rise to decoherence-free subspaces and long-lived quantum coherence. The total Hamiltonian of the system and the common reservoir is given
by

\begin{equation}
H=
\frac{\omega_0}{2}
\sum_{i=1}^{3}
\sigma_z^{(i)}
+
\sum_k
\omega_k
b_k^{\dagger}b_k
+
S_z
\left(
B+B^{\dagger}
\right),
\end{equation}

\noindent where

\begin{equation}
S_z
=
\sum_{i=1}^{3}
\sigma_z^{(i)}
\end{equation}

\noindent is the collective spin operator of the three-qubit system and

\begin{equation}
B
=
\sum_k
g_k b_k
\end{equation}

\noindent is the environmental operator of the common reservoir.\\

\noindent The first term of the Hamiltonian describes the free evolution of three qubits, the second term represents the free Hamiltonian of the bosonic reservoir, and the third term accounts for the collective interaction between the qubits and the common environment. As in the local dephasing model, the interaction Hamiltonian commutes with the system Hamiltonian,

\[
[H_S,H_{SB}]=0,
\]

\noindent and therefore no exchange of energy takes place between the system and the reservoir. The populations of the computational basis states remain constant, whereas the phase relations evolve non-trivially due to the action of the common environment.\\

\noindent A distinctive feature of the collective interaction is that all three-qubits experience identical environmental fluctuations. Consequently, the dephasing process becomes correlated and may generate collective protection mechanisms. In particular, certain multipartite states occupy decoherence-free subspaces in which the environmental action is
effectively cancelled, thereby preserving quantum coherence for extended periods of time. Under the assumption of an initially uncorrelated system-environment state and within the weak-coupling approximation, the reduced density operator obeys the master equation

\begin{equation}
\frac{d\rho(t)}{dt}
=
-i[H_S,\rho(t)]
+
\gamma(t)
S_z\rho(t)S_z
-
\alpha(t)
S_z^2\rho(t)
-
\alpha^*(t)
\rho(t)S_z^2,
\end{equation}

\noindent where the time-dependent coefficients $\gamma(t)$ and $\alpha(t)$ are completely determined by the spectral properties of the common reservoir.\\

\noindent The common environment is assumed to possess an Ohmic spectral density,

\begin{equation}
J(\omega)
=
\eta
\omega
\exp
\left(
-\frac{\omega}{\Lambda}
\right),
\end{equation}

\noindent where $\eta$ denotes the collective system-environment coupling strength and $\Lambda$ represents the cut-off frequency. When the bath correlation time is negligibly small, the environmental memory effects disappear and information flows irreversibly from the system to the reservoir. The resulting dynamics becomes Markovian and coherence decays monotonically.\\

\noindent On the other hand, finite bath correlation times give rise to non-Markovian dynamics. In this regime, the common reservoir possesses memory and can temporarily return information previously lost by the system. The information backflow may substantially suppress coherence degradation, generate long lived coherence, and even produce decoherence-free behaviour for suitable multipartite states. The local and common dephasing models introduced above constitute the theoretical framework employed in the subsequent sections to investigate the dynamics of basis-independent quantum coherence of various tripartite pure and mixed states in both Markovian and non-Markovian regimes.

%====================================================
\section{basis-independent Measure of Quantum Coherence}
%====================================================

\noindent Quantum coherence originates from the superposition principle and constitutes one of the most fundamental nonclassical
features of quantum mechanics. Within the resource-theoretic framework, several coherence quantifiers have been proposed, among which the relative entropy of coherence and the $l_{1}$-norm of coherence are the most widely employed \cite{hz2018,srana2017,zhu2018}. However, these measures are inherently basis-dependent since their values are defined with respect to a prescribed reference basis. Consequently, the amount of coherence assigned to a quantum state may change under a unitary transformation of the basis even though the physical state itself remains unchanged. To overcome this limitation, Radhakrishnan \textit{et al.} introduced a basis-independent measure of quantum coherence constructed from the quantum Jensen--Shannon divergence between a quantum state and the maximally mixed state \cite{Radhakrishnan2019}. Since the maximally mixed state is invariant under arbitrary unitary transformations, the resulting coherence measure provides an intrinsic characterization of the coherence contained in a quantum state and is therefore independent of any particular representation.\\

\noindent For a density operator $\rho$ acting on a
$d$-dimensional Hilbert space, the basis-independent
coherence is defined as

\begin{equation}
C_{BI}(\rho)
=
\sqrt{
S\left(
\frac{\rho+\rho_{m}}{2}
\right)
-
\frac{
S(\rho)
+
S(\rho_{m})
}{2}
},
\label{eq:BIC}
\end{equation}

\noindent where

\begin{equation}
\rho_{m}
=
\frac{\mathbb{I}_{d}}{d}
\label{eq:maxmixed}
\end{equation}

\noindent denotes the maximally mixed state and

\begin{equation}
S(\rho)
=
-\mathrm{Tr}
\left(
\rho \log_{2}\rho
\right)
\label{eq:vnentropy}
\end{equation}

\noindent is the von Neumann entropy. Equation~(\ref{eq:BIC}) represents the square root of the
quantum Jensen-Shannon divergence between the state $\rho$ and the maximally mixed state $\rho_{m}$. Since both the von Neumann entropy and the maximally mixed state are invariant under unitary transformations, the coherence measure (Eq.~(\ref{eq:BIC})) satisfies

\begin{eqnarray}
C_{BI}
\left(
U\rho U^{\dagger}
\right)
=
C_{BI}(\rho),
\label{eq:unitary}
\end{eqnarray}

\noindent for every unitary operator $U$. Consequently, the measure quantifies the intrinsic coherence of a quantum state rather than the coherence associated with a particular choice of basis. In the present work, we are concerned with three-qubit (or tripartite) systems for which the Hilbert space dimension is $d=8$. Accordingly, the maximally mixed state is given by

\begin{eqnarray}
\rho_{m}
=
\frac{\mathbb{I}_{8}}{8},
\label{eq:maxmixed8}
\end{eqnarray}

\noindent and its entropy assumes the value

\begin{equation}
S(\rho_{m})
=
\log_{2}8
=
3.
\label{eq:Sm}
\end{equation}

\noindent Therefore, the basis-independent coherence employed throughout the present investigation can be expressed as

\begin{equation}
C_{BI}(\rho)
=
\sqrt{
S\left(
\frac{\rho+\frac{\mathbb{I}_{8}}{8}}{2}
\right)
-
\frac{
S(\rho)+3
}{2}
}.
\label{eq:BIC3qubit}
\end{equation}

\noindent The quantity defined in Eq.~(\ref{eq:BIC3qubit}) will be used to investigate the dynamical behaviour of intrinsic quantum coherence for various classes of three-qubit pure and mixed states subjected to local and common dephasing environments in both Markovian and non-Markovian regimes.

\subsection{Physical origin of the robustness of basis-independent coherence}

\noindent The remarkable robustness observed throughout the present work can be understood directly from the mathematical structure of the basis-independent coherence measure. Unlike the relative entropy of coherence, which depends explicitly on the off-diagonal elements of the density matrix in a prescribed basis, the basis-independent coherence depends only on the spectrum of the density operator through the quantum Jensen-Shannon divergence with respect to the maximally mixed state.\\

\noindent For an arbitrary density operator $\rho(t)$, the basis-independent coherence $C_{BI}(\rho(t))$ (in a $d$ dimensional system) is

\begin{eqnarray}
\label{eq:cbicbi}
C_{BI}(\rho(t))
=
\sqrt{
S\!\left(
\frac{\rho(t)+I_d/d}{2}
\right)
-
\frac{
S(\rho(t))
+
\log_2 d
}{2}
}.
\end{eqnarray}

\noindent Since the maximally mixed state ($\frac{I_d}{d}$) is invariant under all unitary transformations, the quantity $C_{BI}(\rho(t))$ of Eq.~(\ref{eq:cbicbi}) (and consequently the Eq.~(\ref{eq:BIC3qubit}) of previous section) depends only on the eigenvalues of $\rho(t)$ and is completely insensitive to the basis in which the density matrix is represented. Consequently, environmental processes that predominantly suppress the off-diagonal elements of the density matrix while inducing only weak modifications of its eigenvalue spectrum are expected to produce relatively small variations of the basis-independent coherence.\\

\noindent Pure dephasing channels constitute precisely such a class of quantum evolutions. During pure dephasing, the populations of the computational basis remain unchanged while only the phase relations decay. Although the off-diagonal elements decay with time, the corresponding changes in the eigenvalue spectrum are generally much less pronounced under pure dephasing than the suppression of the coherences themselves. Since the von Neumann entropy depends exclusively on the eigenvalues of $\rho(t)$, both $S(\rho(t))$ and $S\Big(\frac{(\rho(t) + \frac{I_d}{d})}{2}\Big)$ remain comparatively stable throughout the evolution.\\

\noindent This spectral stability provides a natural explanation for the numerical results obtained in the subsequent sections. Under Markovian local dephasing, the gradual modification of the eigenvalue distribution produces a slow reduction of $C_{BI}(\rho(t))$. Collective dephasing alters the spectrum even less owing to the existence of decoherence-free sectors, thereby preserving a larger amount of intrinsic coherence. In the non-Markovian regime, the information backflow from the environment continuously compensates for the small spectral changes induced by dephasing. Consequently, the eigenvalue distribution remains almost unchanged and the basis-independent coherence exhibits the nearly frozen behaviour observed in \textcolor{blue}{Figs.~$2$-$5$} (in the subsequent sections).\\

\noindent The above argument also explains why the basis-independent coherence behaves qualitatively differently from the relative entropy of coherence. The latter directly measures the distance between $\rho$ and its diagonal part and therefore responds immediately to the decay of the off-diagonal matrix elements. In contrast, the basis-independent coherence probes the eigenvalue spectrum of the quantum state through the quantum Jensen-Shannon divergence. Consequently, under the pure dephasing dynamics considered in the present work, it is considerably less sensitive to phase damping and exhibits significantly greater robustness than the relative entropy of coherence.

\section{Pure Tripartite States:}

\noindent In this section, we investigate the dynamics of basis-independent quantum coherence for several representative three-qubit pure states subjected to pure dephasing. The evolution of intrinsic coherence is analyzed under both local and common dephasing environments in the Markovian as well as non-Markovian regimes. To this end, we consider four important classes of tripartite states, namely the Greenberger-Horne-Zeilinger ($|GHZ\rangle$), $|W\rangle$, $|W\overline{W}\rangle$, and $|Star\rangle$ states. Before presenting the coherence dynamics, we briefly discuss the essential features of these states and their correlation structures. Multipartite pure states of three-qubits are broadly classified into two inequivalent categories under stochastic local operations and classical communication (SLOCC), namely the $|GHZ\rangle$ class and the $|W\rangle$ class. These two classes represent distinct forms of genuine multipartite entanglement and cannot be transformed into one another by local operations assisted by classical communication \cite{coffman2000}.\\

\noindent The $|GHZ\rangle$ state is a maximally entangled tripartite state and is defined as

\begin{eqnarray}
|GHZ\rangle
=
\frac{1}{\sqrt{2}}
\left(
|000\rangle
+
|111\rangle
\right),
\label{eq:GHZ}
\end{eqnarray}

\noindent A remarkable property of the GHZ state is that its entanglement is entirely global in nature. Consequently, the loss of any one of its constituent qubits destroys all quantum correlations among the remaining qubits.

\noindent The $|W\rangle$ state, which belongs to a different SLOCC class, is given by

\begin{eqnarray}
|W\rangle
=
\frac{1}{\sqrt{3}}
\left(
|100\rangle
+
|010\rangle
+
|001\rangle
\right).
\label{eq:W}
\end{eqnarray}

\noindent Unlike the $|GHZ\rangle$ state, the entanglement present in the $|W\rangle$ state is distributed in a more robust manner. The bipartite correlations survive even after the loss of one qubit, implying that the state retains a finite amount of quantum correlation under partial decoherence. Besides the $|GHZ\rangle$ and $|W\rangle$ states, we also consider two other physically significant tripartite states, namely the $|W\overline{W}\rangle$ state and the $|Star\rangle$ state. The $|W\overline{W}\rangle$ state is defined as an equal superposition of the $|W\rangle$ state and its spin-flipped counterpart,

\begin{equation}
|W\overline{W}\rangle
=
\frac{1}{\sqrt{2}}
\left(
|W\rangle
+
|\overline{W}\rangle
\right).
\label{eq:WWbar}
\end{equation}

where

\begin{equation}
|\overline{W}\rangle
=
\frac{1}{\sqrt{3}}
\left(
|011\rangle
+
|101\rangle
+
|110\rangle
\right).
\label{eq:Wbar}
\end{equation}

\noindent The $W\overline{W}$ state possesses a rich correlation structure. In particular, quantum coherence and correlations are simultaneously present at the single qubit, bipartite, and tripartite levels. Owing to the coexistence of quantum features at different hierarchical levels, this state provides an excellent platform for investigating the distribution and robustness of multipartite quantum resources. The $|GHZ\rangle$, $|W\rangle$, and $|W\overline{W}\rangle$ states are permutation symmetric. Therefore, the reduced bipartite states obtained by tracing over any one of the qubits are equivalent and exhibit identical coherence and entanglement properties irrespective of the choice of the discarded qubit.

\noindent In contrast, the $|Star\rangle$ state is intrinsically asymmetric and is defined as

\begin{equation}
|Star\rangle
=
\frac{1}{2}
\left(
|000\rangle
+
|100\rangle
+
|101\rangle
+
|111\rangle
\right).
\label{eq:Star}
\end{equation}

\noindent Similar to the $|W\overline{W}\rangle$ state, the $|Star\rangle$ state exhibits quantum coherence and correlations at different levels of the multipartite system. However, these correlations are distributed non-uniformly among the qubits. The first and second qubits play the role of peripheral qubits, whereas the third qubit acts as a central qubit. Consequently, tracing over the central qubit renders the remaining qubits separable, while tracing over either of the peripheral qubits still leaves a finite amount of bipartite entanglement \cite{roy2023}. In the following subsections, we employ the basis-independent coherence measure introduced in \textcolor{blue}{Sec.~$2$} to investigate the temporal evolution of intrinsic coherence for the four tripartite states defined in Eqs.~(\ref{eq:GHZ})-(\ref{eq:Star}) under local and common dephasing environments in both Markovian and non-Markovian regimes.\\

%\noindent  For convenience, the coherence dynamics are presented as functions of the evolution time $t$.

\subsection{Pure States Subjected to Markov and non-Markov Dephasing Noise}
\begin{figure}
    \centering
    \includegraphics[width=0.8\linewidth]{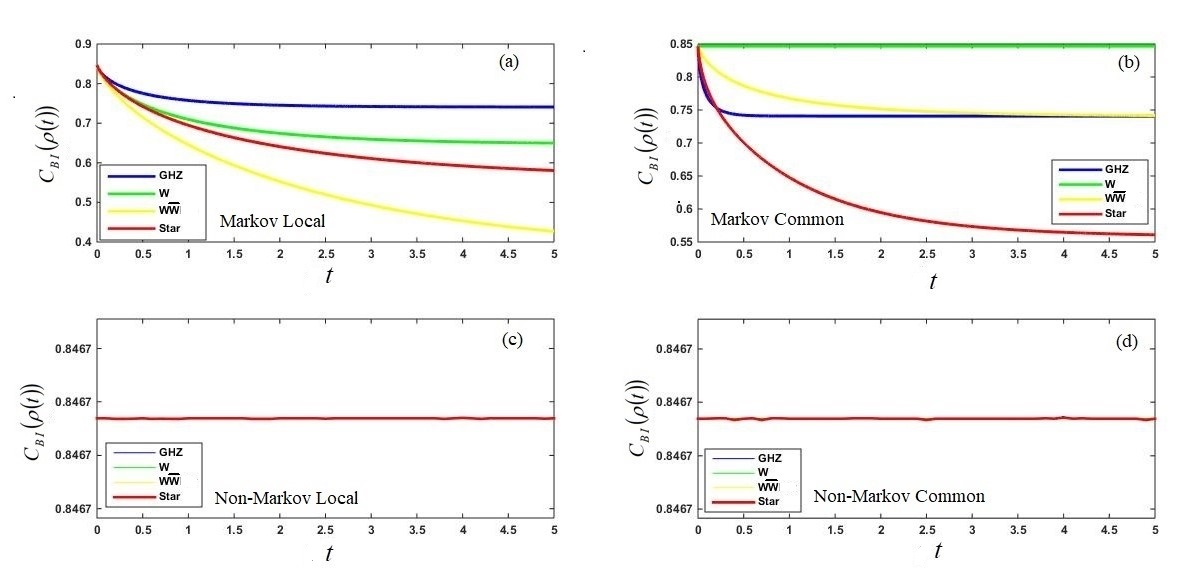}
    \caption{The figure shows the temporal evolution of the basis-independent coherence $C_{BI}(\rho(t))$ for the four tripartite pure states, namely the $|GHZ\rangle$, $|W\rangle$, $|W\overline{W}\rangle$, and $|Star\rangle$ states under local and common dephasing environments in both Markovian and non-Markovian regimes.}
    \label{fig:placeholder}
\end{figure}

.
\begin{center}
    \textit{Markovian local environment} 
\end{center}

\noindent \textcolor{blue}{Fig.~$2(a)$} depicts the Markovian local dephasing model. All the four pure states show the same amount of intrinsic coherence initially, nearly $0.85$. As the interaction with the independent local reservoirs ($L_{E_1}, L_{E_2}$ and $L_{E_3}$) progresses, monotonic decay of coherence $(C_{BI}(\rho(t))$ is observed for all the four pure states ($|GHZ\rangle, |W\rangle, |W\overline{W}\rangle$ and $|Star\rangle$) and this is due to the irreversible loss of phase information. However, the decay rates are strongly state dependent. $|GHZ\rangle$ state exhibits the highest robustness among the four pure states and approaches a finite asymptotic value of coherence. For $t>~1.5$, there is almost no decay in the amount of coherence of the $|GHZ\rangle$ state. The $|W\rangle$ and $|Star\rangle$ states display an intermediate behaviour in decay of coherence. For $|W\rangle$ state coherence is preserved when $t> 3.5$ (approx.) while in this region that is for $t> 3.5$, the coherence of $|Star\rangle$ state further decays. On the other hand, $|W\overline{W}\rangle$ state suffers the most reduction in the amount of coherence and reaches to the smallest asymptotic value. The monotonic decrease of the amount of coherence of all the curves, shown in \textcolor{blue}{Fig.~$2(a)$} is a direct consequence of the memoryless nature of the Markovian local environment in the case of the four pure states. This indicates that when these pure states are subjected to Markovian local dephasing, the $|GHZ\rangle$ state is less affected while the $|W\overline{W}\rangle$ state is the most affected, indicating how the pure states, when used as quantum channel for information transfer, are prone to the effects of dephasing environment.\\

\begin{center}
    \textit{Markovian common environment} 
\end{center}

\noindent As can be seen from \textcolor{blue}{Fig.~$2(b)$}, the behaviour with respect to the the decay in the amount of coherence changes significantly when the pure states $|GHZ\rangle, |W\rangle, |W\overline{W}\rangle$ and $|Star\rangle$ are subjected to Markovian common environment ($C_{E}$). The $|W\rangle$ state exhibits an almost constant coherence throughout the evolution and is practically immune to the collective dephasing process. There is moderate reduction in the amount of coherence $(C_{BI}(\rho(t))$ for $|GHZ\rangle$ and $|W\overline{W}\rangle$ states and subsequently these coherences attain stationary values with increase in $t$. It is to be noted from the \textcolor{blue}{Fig.~$2(b)$} that from the initial value there is steep decrease in $(C_{BI}(\rho(t))$ for $|GHZ\rangle$ state and then retains coherence value of $0.75$ (aprrox.) for $t > 0.5$, whereas the decaying of coherence of $|W\overline{W}\rangle$ state from its initial value is slow in comparison to the $|GHZ\rangle$ state and attains $0.75$ approx. when $t > 2.5$. However, the coherences of these two state (i.e. $|GHZ\rangle$ and $|W\overline{W}\rangle$) coincide with one another for $t > 3$. In contrast, the $|Star\rangle$ state experiences a comparatively rapid decrease in the amount of coherence and stabilizes at a considerably lower value of coherence with respect to the other three pure states (i.e. $|GHZ\rangle$, $|W\overline{W}\rangle$ and $|W\rangle$). The enhanced preservation of coherence in the common Markovian environment can be attributed to the correlated action of the reservoir on all three-qubits, which permits the existence of decoherence free subspaces. The pure states having support predominantly within these protected subspaces retain a substantial amount of their intrinsic coherence under common Markovian dephasing environment ($C_E$).  \\

\begin{center}
    \textit{non-Markovian local and common environment} 
\end{center}

\noindent \textcolor{blue}{Figs.~$2(c)$ and $2(d)$} respectively depict the coherence dynamics in the non-Markovian local and non-Markovian common dephasing environments. Remarkably, the basis-independent coherence remains essentially constant during the entire evolution and the curves for all four pure states overlap almost perfectly. The coherence assumes a value close to $C_{BI}(\rho(t))\approx0.8467$ and exhibits only negligible fluctuations. The fluctuations in non-Markovian common environment are little more than those observed in non-Markov local environment (\textcolor{blue}{See  figs.~$2(c)$ and $2(d)$}). The nearly frozen behaviour observed here is consistent with the general mechanism discussed in \textcolor{blue}{Sec.~$III.1$}. Since the basis-independent coherence is determined by entropy functionals of the eigenvalue spectrum of the density operator, the weak spectral modifications induced by non-Markovian pure dephasing lead to only negligible temporal variations of the coherence.\\

\noindent Overall, \textcolor{blue}{Fig.~$2$} demonstrates that in particular, the non-Markovian environments preserve the intrinsic coherence of all considered tripartite pure states almost perfectly, whereas the Markovian regimes exhibit state dependent degradation with the common environment providing considerably better protection than the local environment (indeed not for the $|Star\rangle$ state).

\section{Mixed Tripartite States:}
\noindent Since the preparation and maintenance of pure quantum states are often hindered by experimental imperfections and environmental interactions, it is imperative to investigate coherence in mixed states to obtain a more complete understanding of quantum systems. Accordingly, we analyze three representative mixed states such as (i) $|GHZ\rangle - |W\rangle$ mixture ($\rho_{GW}$), (ii) $|Werner\rangle - |GHZ\rangle$ mixture ($\rho_{WrG}$)  and (iii) $|Werner\rangle - |W\rangle$ mixture ($\rho_{WrW}$). In all the three mixtures the mixing parameter is chosen to be $p$ and the values $p=0.1$, $0.5$, and $0.9$ are chosen as representatives of low, intermediate, and high mixing regimes, respectively. These values avoid the trivial limiting cases corresponding to pure states and enable a clear illustration of the dependence of coherence dynamics on the degree of mixing across the entire parameter range (i.e. \(0 \leq p \leq 1\)). The three mixed states collectively provide a comprehensive framework for investigating the behaviour of intrinsic multipartite coherence under both structural and statistical imperfections. They allow us to probe the interplay between different entanglement structures, environmental decoherence, and classical noise within a unified setting.\\

%\noindent As before, the coherence dynamics are presented as functions of the evolution time t..

\subsection{ Mixture of $|GHZ\rangle$ and $|W\rangle$ state}
\noindent The mixture of the  $|GHZ\rangle$ and $|W\rangle$ state is defined as
\begin{equation}
\rho_{GW}
=
p\,|\mathrm{GHZ}\rangle\langle \mathrm{GHZ}|
+
(1-p)\,|W\rangle\langle W|,
\label{eq:GW}
\end{equation}
where \(p\) denotes the mixing probability (with \(0 \leq p \leq 1\)). This state provides an interpolation between the two inequivalent classes of genuine tripartite entanglement, namely the $|GHZ\rangle$ and $W\rangle$ classes, which possess markedly different robustness properties under decoherence. The study of this mixture enables us to understand the competition between fragile and robust forms of multipartite quantum correlations.\\

\noindent Since $\langle GHZ|W\rangle=0$, the two states form an orthogonal pair and the eigenvalues of $\rho_{GW}$ are $\{p,1-p,0,0,0,0,0,0\}$. Thus, the spectrum is symmetric under the transformation $p\rightarrow1-p$, and consequently the basis-independent coherence satisfies $C_{\rm BI}(p)=C_{\rm BI}(1-p)$. At $p=0.5$, the two orthogonal components have equal statistical weights, resulting in the maximum entropy associated with this two coherent component spectrum and hence the minimum basis-independent coherence. Importantly, the mixing is statistical and does not introduce coherence between $|GHZ\rangle$ and $|W\rangle$. Consequently, the equal mixture does not inherit the full intrinsic coherence of either pure state. As $p$ moves away from $0.5$, one of the two coherent components becomes dominant and the state approaches a single pure state in the limits $p\rightarrow0$ or $p\rightarrow1$. Therefore, the basis-independent coherence increases again toward the two pure-state limits, producing a minimum at the equal mixture.

\subsubsection{$|GHZ\rangle-|W\rangle$ Mixture Subjected to Markov and non-Markov Dephasing Noise}

\begin{center}
    \textit{Markovian local environment} 
\end{center}

\begin{figure}
    \centering
    \includegraphics[width=0.8\linewidth]{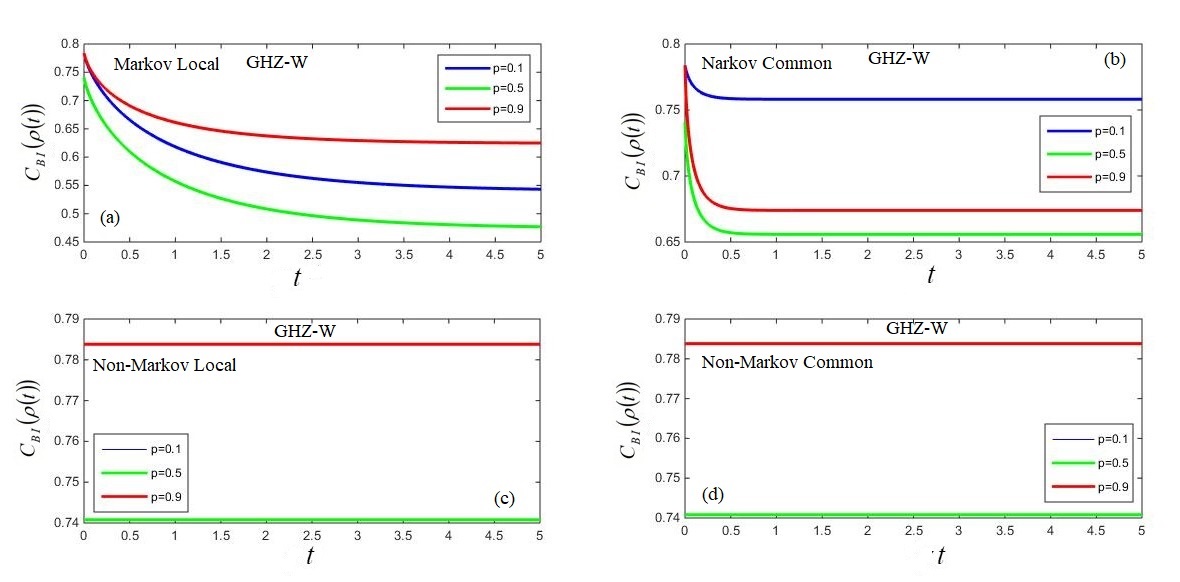}
    \caption{The figure shows the temporal evolution of the basis-independent coherence $C_{BI}(\rho(t))$ for the $|GHZ\rangle-|W\rangle$ mixture (i.e. of $\rho_{GW}$) under local and common dephasing environments in both Markovian and non-Markovian regimes for different values of the mixing probability $p$.}
    \label{fig:placeholder}
\end{figure}

\noindent \textcolor{blue}{Fig.$3(a)$} depicts the evolution of basis-independent coherence ($C_{BI}(\rho(t)$) under Markovian local dephasing environment. Initially, for mixing parameters $p=0.9$ and $p=0.1$ possess nearly the same amount of intrinsic coherence with $C_{BI}(\rho(t)) = 0.78$ (approx.). For $p=0.5$, however, the intrinsic initial coherence starts nearly at $0.75$. As the interaction with the independent reservoir proceeds, the coherence decreases monotonically and eventually approaches non-vanishing stationary values. It is also to be noted that the decay behaviour depends significantly on the mixing parameter $p$. Among the considered cases, the state with the equal-weight mixture ($p=0.5$) shows the most pronounced reduction in the amount of coherence and reaches to the smallest asymptotic value, while the state $\rho_{GW}$ with high mixing $p = 0.9$ retains the largest residual coherence. For low mixing at $p = 0.1$, the state $\rho_{GW}$ displays an intermediate behaviour with respect to the decaying in the amount of coherence. For $t> 4$, the amount of coherence for each value of mixing parameter stabilizes to certain values.\\

\begin{center}
    \textit{Markovian common environment} 
\end{center}

\noindent The behaviour changes appreciably in the Markovian common dephasing environment shown in Fig.~$3(b)$. In this case, the basis-independent coherence $C_{BI}(\rho(t))$ undergoes a rapid initial decrease followed by a saturation to stationary values within a short time. Interestingly, the mixture $\rho_{GW}$ with $p=0.1$ (low mixing) preserves the largest amount of intrinsic coherence throughout the evolution, whereas the state $\rho_{GW}$ with $p=0.5$ (moderate mixing) attains the smallest stationary value. The state $\rho_{GW}$ with $p=0.9$ (high mixing) exhibits an intermediate degree of robustness. The enhanced preservation of coherence in the common Markovian environment can be attributed to the correlated interaction of three qubits with a single reservoir $C_E$. In particular, the presence of the $|W\rangle$ component, which occupies a decoherence free sector of the collective dephasing dynamics, provides substantial protection against coherence loss and results in finite residual coherence even at long times. 

\begin{center}
    \textit{non-Markovian local and common environment} 
\end{center}

\noindent \textcolor{blue}{Figs.~ $3(c)$ and $3(d)$} depict the coherence dynamics in the non-Markovian local and non-Markovian common environments, respectively. Remarkably, the basis-independent coherence remains essentially constant during the entire evolution and the curves exhibit negligible temporal variation. The coherence assumes nearly constant values of approximately $C_{BI}(\rho(t))\approx0.74$ for $p=0.1$ (low mixing) and $p=0.5$ (moderate mixing), and $C_{BI}(\rho(t))\approx0.784$ for $p=0.9$ (high mixing). The overlap of the curves for $p=0.1$ and $p=0.5$ indicates that these two mixtures possess almost identical intrinsic coherence under non-Markovian evolution. The persistence of nearly frozen coherence follows the same spectral mechanism discussed in \textcolor{blue}{Sec.~$III.1$}. Although the mixing parameter $p$ changes the initial eigenvalue distribution, the non-Markovian evolution introduces only weak spectral modifications, resulting in an almost invariant basis-independent coherence.\\

\noindent Overall, \textcolor{blue}{Fig.~$3$} reveals that the basis-independent coherence of the $|GHZ\rangle-|W\rangle$ mixture is remarkably robust against dephasing noise. While the Markovian regimes exhibit a nontrivial dependence on the mixing probability and environmental configuration, the non-Markovian environments preserve the intrinsic coherence almost perfectly. Furthermore, the collective interaction provides significantly better protection than the independent local reservoirs owing to the emergence of decoherence-free behaviour associated with the $|W\rangle$ component of the mixed state.

\subsection{ $|Werner\rangle - |GHZ\rangle$ Mixture}
\noindent The $|Werner\rangle - |GHZ\rangle$ mixture is a mixed state, which is defined as
\begin{equation}
\rho_{WrG}
= p\, |\mathrm{GHZ}\rangle\langle \mathrm{GHZ}|
+ \frac{1-p}{8}\, I_{8},
\label{eq:WernerGHZ}
\end{equation}
where \(p\) is the probability of mixing (or mixing parameter) and \(I_{8}\) is the identity matrix of order \(8\). The state in Eq.~\eqref{eq:WernerGHZ} is a regular tripartite mixed state obtained as a convex combination of the $|GHZ\rangle$ state and the maximally mixed state \(\frac{I_8}{8}\) (the terminology $|Werner\rangle - |GHZ\rangle$ mixture comes from the definition of Werner state, which in bipartite system is a convex mixture of pure state and maximally mixed state). This state represents a $|GHZ\rangle$ state contaminated by isotropic white noise and serves as a realistic model of imperfect state preparation. It allows us to investigate how the intrinsic coherence of a highly fragile tripartite state survives in the presence of statistical noise.

\subsubsection{$|Werner\rangle - |GHZ\rangle$ Mixture Subjected to Markov and non-Markov Dephasing Noise}
\begin{figure}
    \centering
    \includegraphics[width=0.8\linewidth]{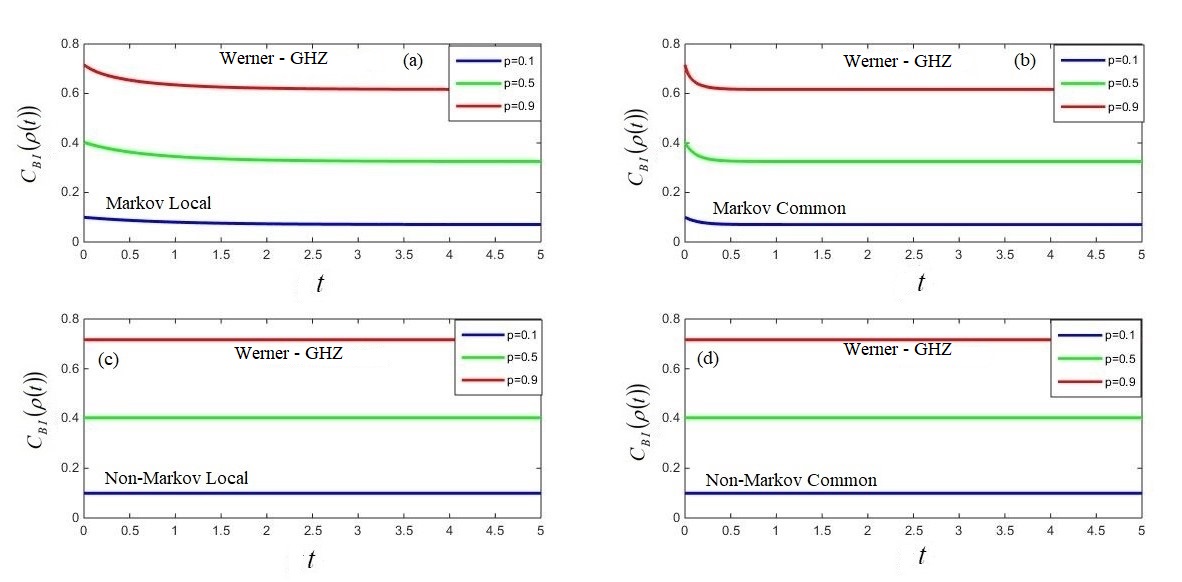}
    \caption{The figure shows the temporal evolution of the basis-independent coherence $C_{BI}(\rho(t))$ for the $|Werner\rangle - |GHZ\rangle$ mixture (i.e. of $\rho_{WrG}$) under local and common dephasing environments in both Markovian and non-Markovian regimes for different values of the mixing probability $p$.}
    \label{fig:placeholder}
\end{figure}

\begin{center}
    \textit{Markovian local environment} 
\end{center}  

\noindent \textcolor{blue}{Fig.~4(a)} shows the evolution of the basis-independent coherence of the $|Werner\rangle-|GHZ\rangle$ mixture under Markovian local dephasing. The basis-independent coherence $C_{BI}(\rho(t))$ of $\rho_{WrG}$ decreases monotonically with the time component $t$ and eventually approaches to non-vanishing stationary values. The stationary values are achieved for $t > 2$ for low, moderate as well as high mixing values of the parameter $p$. The state with $p = 0.9$ exhibits the largest intrinsic coherence throughout the evolution and stabilizes at the highest asymptotic value. The state $\rho_{WrG}$ corresponding to the moderate mixing $p = 0.5$ shows intermediate behaviour, whereas the mixture with $p=0.1$ (low mixing) possesses the smallest coherence and rapidly approaches to a stationary value. The monotonic reduction of coherence originates from the irreversible flow of information from the system to the independent reservoirs, which continuously suppresses the phase correlations associated with the $|GHZ\rangle$ component of the state.\\

\begin{center}
    \textit{Markovian common environment} 
\end{center}

\noindent The behaviour in the Markovian common environment shown in \textcolor{blue}{Fig.~4(b)} is qualitatively similar, although the initial decay becomes considerably faster and the stationary regime is reached within a shorter time interval. Nevertheless, the residual coherence values remain ordered according to the mixing parameter, with
\[
C_{BI}^{(p=0.9)}(\rho_{WrG})
>
C_{BI}^{(p=0.5)}(\rho_{WrG})
>
C_{BI}^{(p=0.1)}(\rho_{WrG}).
\]
\noindent The rapid saturation is a consequence of the correlated action of the common environment on three qubits. Since the maximally mixed component is invariant under dephasing, the long-time coherence is predominantly determined by the surviving contribution of the $|GHZ\rangle$ component and therefore increases with increasing values of $p$.

\begin{center}
    \textit{non-Markovian local and common environment} 
\end{center}

\noindent  \textcolor{blue}{Figs.~ $4(c)$ and $4(d)$} depict the coherence dynamics in the non-Markovian local and non-Markovian common environments, respectively. Remarkably, the basis-independent coherence remains practically constant during the entire evolution and the curves exhibit negligible temporal variations. The coherence assumes approximately constant values of $C_{BI}(\rho_{WrG})\approx0.10$, $0.40$, and $0.70$ for $p=0.1$, $0.5$, and $0.9$, respectively. In contrast to the $|GHZ\rangle - |W\rangle$ mixture in the non-Markovian local and common regimes (\textcolor{blue}{shown in Fig.~$3(c)$ and $3(d)$}), the three curves remain clearly separated and do not overlap, indicating that the intrinsic coherence of the $|Werner\rangle - |GHZ\rangle$ state depends strongly on the amount of the $|GHZ\rangle$ component present in the initial state. The nearly frozen behaviour again follows from the spectral robustness discussed in \textcolor{blue}{Sec.~III.1}. Since the eigenvalue distribution changes only weakly under non-Markovian dephasing, the entropy based basis-independent coherence remains almost constant throughout the evolution.\\

\noindent Overall, \textcolor{blue}{Fig.~$4$} reveals that the basis-independent coherence of the $|Werner\rangle - |GHZ\rangle$ state exhibits a pronounced dependence on the mixing parameter $p$. Increasing the weight of the $|GHZ\rangle$ component significantly enhances the amount of intrinsic coherence preserved by the system. Furthermore, the non-Markovian environments almost perfectly protect the intrinsic coherence, whereas the Markovian regimes induce a monotonic but incomplete degradation of coherence. The persistence of finite stationary values in all cases originates from the presence of the maximally mixed component, which remains unchanged under dephasing and prevents the complete disappearance of the basis-independent coherence.

\subsection{$|Werner\rangle - |W\rangle$ Mixture}

\noindent The next mixed state that we consider here is the $|Werner\rangle - |W\rangle$ mixture, which is defined as
\begin{equation}
\rho_{WrW}
= p\, |W\rangle\langle W|
+ \frac{1-p}{8}\, I_{8},
\label{eq:WernerW}
\end{equation}
where \(p\) is the probability (mixing parameter) and \(I_{8}\) is the identity matrix of order \(8\). The state in Eq.~\eqref{eq:WernerW} is a regular tripartite mixed state obtained as a convex combination of the \(W\) state and the maximally mixed state \(\frac{I_8}{8}\). This state describes a robust tripartite coherent state mixed with white noise. The study of this state enables us to examine whether the well known robustness and decoherence free characteristics of the $|W\rangle$ state persist even in the presence of environmental and preparation induced imperfections. 

\subsubsection{$|Werner\rangle - |W\rangle$ Mixture Subjected to Markov and non-Markov Dephasing Noise}
\begin{figure}
    \centering
    \includegraphics[width=0.8\linewidth]{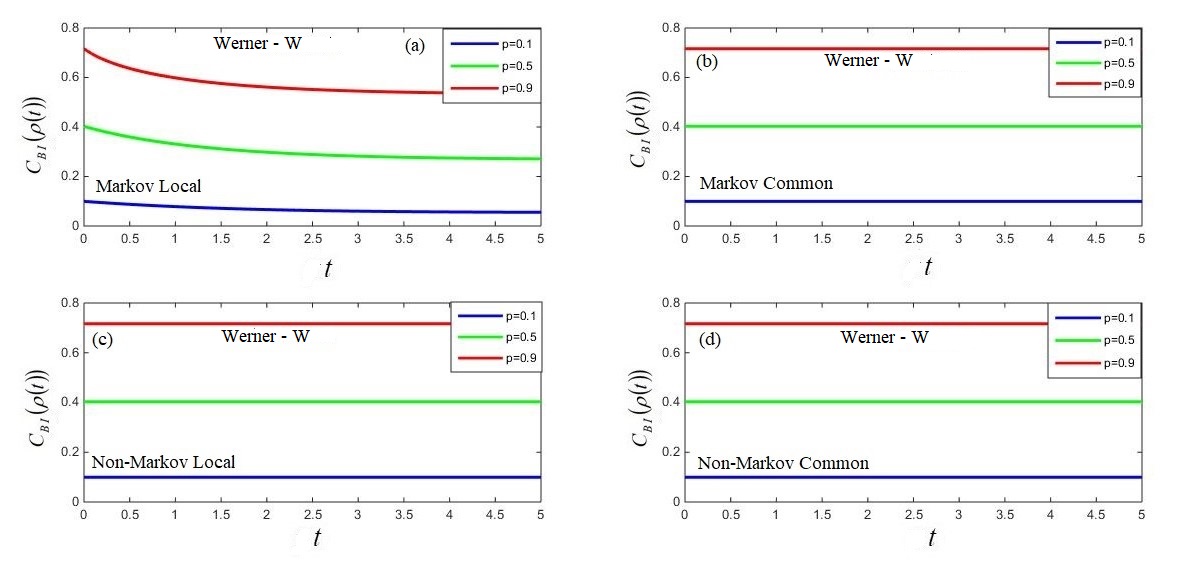}
    \caption{The figure shows the temporal evolution of the basis-independent coherence $C_{BI}(\rho(t))$ for the $|Werner\rangle - |W\rangle$ mixture (i.e. of $\rho_{WrW}$) under local and common dephasing environments in both Markovian and non-Markovian regimes for different values of the mixing probability $p$.}
    \label{fig:placeholder}
\end{figure}

\begin{center}
    \textit{Markovian local environment} 
\end{center}

\noindent \textcolor{blue}{Fig.~ $5(a)$} shows the Markovian local dephasing environment with respect to the $|Werner\rangle - |W\rangle$ mixture. Initially, the intrinsic coherence depends strongly on the mixing parameter $p$ and increases with increasing values of $p$. As a result to the interaction with the independent reservoirs ($L_{E_1},~L_{E_2},~L_{E_3}$), the coherence decreases monotonically and eventually approaches finite stationary values. The state with $p=0.9$ (high mixing) retains the largest amount of intrinsic coherence throughout the evolution process, while the state with $p=0.1$ (low mixing) possesses the smallest coherence and rapidly stabilizes at a comparatively low value. The mixture corresponding to $p=0.5$ (moderate mixing) exhibits an intermediate behaviour. Although coherence degradation is unavoidable in the memoryless (Markov) local environment, the decay remains stagnant and significant residual coherence survives even at long times. This behaviour reflects the relatively robust distribution of coherence associated with the $|W\rangle$ component of the mixed state.\\

\begin{center}
    \textit{Markovian common environment} 
\end{center}

\noindent The dynamics changes dramatically in the Markovian common environment shown in \textcolor{blue}{Fig.~$5(b)$}. Remarkably, the basis-independent coherence remains essentially constant during the entire evolution and the curves corresponding to all three values of mixing parameter $p$ exhibit negligible temporal variation. The coherence assumes approximately constant values of $C_{BI}(\rho_{WrW})\approx0.10$, $0.40$, and $0.72$ for $p=0.1$, $0.5$, and $0.9$, respectively. The complete suppression of coherence decay indicates that the $|Werner\rangle - |W\rangle$ state is practically immune to collective dephasing under Markovian common environment. This remarkable robustness originates from the presence of the $|W\rangle$ component, which predominantly occupies a decoherence free sector of the collective dynamics and therefore remains effectively protected against phase damping.\\

\begin{center}
    \textit{non-Markovian local and common environment} 
\end{center}

\noindent \textcolor{blue}{Figs.~$5(c)$ and $5(d)$} depict the coherence dynamics in the non-Markovian local and non-Markovian common environments, respectively. Similar to the Markovian common case, the basis-independent coherence remains almost perfectly frozen throughout the evolution. The curves are completely flat and maintain the same coherence values as their initial states, indicating the absence of any coherence degradation. This behaviour is likewise explained by the spectral stability of the reduced density operator discussed in \textcolor{blue}{Sec.~$III.1$}. Consequently, the basis-independent coherence remains almost invariant despite the presence of environmental interactions. Consequently, the intrinsic coherence measured by the basis-independent quantifier becomes extremely robust and remains largely unaffected by both local and collective non-Markovian dephasing environments.\\

\noindent Overall, \textcolor{blue}{Fig.~$5$} reveals that the basis-independent coherence of the $|Werner\rangle - |W\rangle$ state exhibits exceptional robustness against dephasing noise. While the Markovian local environment induces a moderate monotonic reduction of coherence, the Markovian common and both non-Markovian environments preserve the intrinsic coherence almost perfectly. Furthermore, increasing the weight of the $|W\rangle$ component systematically enhances the amount of coherence retained by the system, demonstrating the superior resilience of the $|W\rangle$ state structure against environmental decoherence.

\subsection*{Summary of the coherence dynamics}
\noindent The quantitative values in Table~$I$ provide a direct comparison of the coherence preservation under the four dephasing environments. For the pure states, the Markovian local environment generally produces a substantial reduction of $C_{\rm BI}$, whereas the Markovian common environment preserves a larger fraction of the initial coherence for the $|W\rangle$ and $|W\overline{W}\rangle$ states. The $|Star\rangle$ state exhibits a comparatively stronger reduction under both Markovian environments. In contrast, the non-Markovian local and common environments preserve the coherence nearly unchanged over the considered evolution interval. For the mixed states, the quantitative values further demonstrate the distinct dependence on the mixing parameter for the $|GHZ\rangle- |W\rangle$ mixture and the Werner-type states, as discussed in Sec.~$V$.

\begin{table}[htbp]
\centering
\label{tab1}
\renewcommand{\arraystretch}{1.15}
\begin{tabular}{lcccc}
\toprule
\textbf{Initial state}
& \textbf{ML}
& \textbf{MC}
& \textbf{NL}
& \textbf{NC} \\
\midrule

$|GHZ\rangle$
& 0.741
& 0.740
& 0.847
& 0.847 \\

$|W\rangle$
& 0.650
& 0.847
& 0.847
& 0.847 \\

$|W\overline{W}\rangle$
& 0.427
& 0.741
& 0.847
& 0.847 \\

$|Star\rangle$
& 0.579
& 0.558
& 0.847
& 0.847 \\

\midrule

$\rho_{GW},\ p=0.1$
& 0.543
& 0.758
& 0.741
& 0.741 \\

$\rho_{GW},\ p=0.5$
& 0.476
& 0.656
& 0.741
& 0.741 \\

$\rho_{GW},\ p=0.9$
& 0.625
& 0.678
& 0.784
& 0.784 \\

\midrule

$\rho_{WrG},\ p=0.1$
& 0.067
& 0.066
& 0.100
& 0.100 \\

$\rho_{WrG},\ p=0.5$
& 0.326
& 0.326
& 0.403
& 0.403 \\

$\rho_{WrG},\ p=0.9$
& 0.614
& 0.615
& 0.716
& 0.716 \\

\midrule

$\rho_{WrW},\ p=0.1$
& 0.055
& 0.099
& 0.100
& 0.100 \\

$\rho_{WrW},\ p=0.5$
& 0.270
& 0.399
& 0.403
& 0.403 \\

$\rho_{WrW},\ p=0.9$
& 0.535
& 0.710
& 0.716
& 0.716 \\

\bottomrule
\end{tabular}
\caption{Quantitative comparison of the basis-independent coherence
$C_{\rm BI}$ for the considered tripartite states under Markovian local
(ML), Markovian common (MC), non-Markovian local (NL), and non-Markovian
common (NC) dephasing environments. The ML and MC columns give the
coherence at the final evolution time $t=5$, while the NL and NC values
represent the approximately preserved coherence over the considered
evolution interval. For the mixed states, the corresponding mixing
parameter $p$ is indicated.}
\end{table}

% ============================================================
% Additional analysis requested by the Reviewer
% ============================================================

\section{Partial exposure to dephasing environment}

\noindent To further examine the robustness of tripartite coherence, we have considered an additional partial exposure scenario in which only two of the three qubits are subjected to local Markovian dephasing, while the third qubit is isolated from the environment. Specifically, qubits $A$ and $B$ are exposed to the local Markovian dephasing environment, whereas qubit $C$ is kept noise-free. For the present analysis, qubit $C$ is chosen as the noise-free qubit. This choice is immaterial for the permutation-symmetric states $|GHZ\rangle$, $|W\rangle$ and $|W\overline{W}\rangle$, whereas for the asymmetric $|Star\rangle$ state the protected-qubit choice can in principle affect the quantitative enhancement. A systematic comparison of inequivalent protection configurations is beyond the scope of the present work. The values at the final simulation time, $t=5$, are compared with those obtained when all three qubits are exposed to local Markovian dephasing. This is shown in the following Table $II$.\\
\begin{table}[htbp]
\centering
\label{tab:partial_noise}
\renewcommand{\arraystretch}{1.15}
\begin{tabular}{lcc}
\toprule
\textbf{Initial state}
& \textbf{Three qubits noisy}
& \textbf{$A,B$ noisy; $C$ noise-free} \\
\midrule
$|GHZ\rangle$  & 0.740928 & 0.741705 \\
$|W\rangle$    & 0.649776 & 0.661497 \\
$|W\overline{W}\rangle$ & 0.426663 & 0.597996 \\
$|Star\rangle$ & 0.580540 & 0.669947 \\
\bottomrule
\end{tabular}
\caption{Comparison of the basis-independent coherence $C_{\rm BI}$ at
$t=5$ for the original Markovian local (ML) configuration, in which all
three qubits are exposed to dephasing, and the partial-noise
configuration, in which qubits $A$ and $B$ are exposed to dephasing
while qubit $C$ is noise-free.}
\end{table}

\noindent As shown in Table~$II$, protecting one qubit from the dephasing environment generally results in a higher residual basis-independent coherence. The enhancement is relatively small for the $|GHZ\rangle$ state, whereas it is more pronounced for the $|W\rangle$, $|W\overline{W}\rangle$, and $|Star\rangle$ states, with the largest increase observed for the $|W\overline{W}\rangle$ state. This behavior reflects the reduced dephasing experienced by the tripartite system when one of the local dephasing channels is removed. \\

\noindent We have also extended this partial-noise analysis to the mixed tripartite states considered in Sec.~V, namely the $|GHZ\rangle-|W\rangle$ mixture, the $|Werner\rangle-|GHZ\rangle$ mixture, and the $|Werner\rangle-|W\rangle$ mixture. For all the considered mixing parameters, $p=0.1$, $0.5$, and $0.9$, the isolation of one qubit leads to an enhancement of the residual basis-independent coherence compared with the corresponding configuration in which all three qubits are exposed to local Markovian (ML) dephasing. The magnitude of this enhancement depends on both the mixing parameter and the structure of the initial mixed state. The effect is comparatively weak for the $|Werner\rangle-|GHZ\rangle$ state, while it is more noticeable for the $|GHZ\rangle -|W\rangle$ and $|Werner\rangle-|W\rangle$ states. Thus, the partial-noise analysis confirms that isolation of one qubit generally improves the preservation of basis-independent coherence for both pure and mixed tripartite states.\\

\begin{table}[htbp]
\centering
\label{tab:partial_mixed}
\renewcommand{\arraystretch}{1.1}
\begin{tabular}{lcccc}
\toprule
\textbf{State Mixtures} & \textbf{$p$} & \textbf{All three noisy}
& \textbf{$A,B$ noisy; $C$ free$ $} & \textbf{Increase} \\
\midrule
$\rho_{GW}$  & 0.1 & 0.543272 & 0.557183 & 0.013911 \\
$\rho_{WrG}$ & 0.1 & 0.070532 & 0.071072 & 0.000540 \\
$\rho_{WrW}$ & 0.1 & 0.055317 & 0.060354 & 0.005037 \\
$\rho_{GW}$  & 0.5 & 0.476890 & 0.490265 & 0.013374 \\
$\rho_{WrG}$ & 0.5 & 0.325493 & 0.326499 & 0.001006 \\
$\rho_{WrW}$ & 0.5 & 0.271226 & 0.283925 & 0.012699 \\
$\rho_{GW}$  & 0.9 & 0.624892 & 0.629276 & 0.004384 \\
$\rho_{WrG}$ & 0.9 & 0.616576 & 0.617446 & 0.000869 \\
$\rho_{WrW}$ & 0.9 & 0.534390 & 0.547183 & 0.012792 \\
\bottomrule
\end{tabular}
\caption{Basis-independent coherence $C_{\rm BI}$ at $t=5$ for the
mixed tripartite states under the original ML configuration and under
partial local Markovian dephasing, with qubits $A$ and $B$ exposed to
noise and qubit $C$ noise-free.}
\end{table}

\noindent In the present work, we restrict the partial-exposure analysis to the Markovian local environment, which is sufficient to demonstrate the effect of partial environmental protection.\\

\section{Conclusion}

\noindent In this work, we have investigated the dynamics of basis-independent quantum coherence for several classes of tripartite pure and mixed states subjected to local and common pure-dephasing environments in both Markovian and non-Markovian regimes. The basis-independent coherence measure based on the quantum Jensen--Shannon divergence was employed to quantify the intrinsic coherence of the system. Since this measure is invariant under arbitrary unitary transformations, it provides a basis-independent characterization of coherence and allows the intrinsic coherence properties of different tripartite states to be compared without choosing a preferred reference basis.\\

\noindent For the pure tripartite states, namely $|GHZ\rangle$, $|W\rangle$, $|WW\rangle$, and $|Star\rangle$, the Markovian local environment produces a clear state-dependent degradation of basis-independent coherence. The $|GHZ\rangle$ state retains the largest residual coherence among the states considered, whereas the $|WW\rangle$ state experiences the strongest reduction. The $|W\rangle$ state exhibits a particularly robust behaviour under Markovian common dephasing, where its coherence is almost completely preserved owing to its association with a decoherence-free sector of the collective dynamics. The $|Star\rangle$ state shows a comparatively stronger sensitivity to collective dephasing because of its asymmetric structure.\\

\noindent A markedly different behaviour is obtained in the non-Markovian environments. Under both non-Markovian local and non-Markovian common dephasing, the basis-independent coherence remains almost perfectly frozen throughout the considered evolution interval. The coherence is therefore largely insensitive to the detailed structure of the initial pure state in the presence of environmental memory. This demonstrates that non-Markovian memory effects can strongly suppress the degradation of intrinsic multipartite coherence.\\

\noindent For the mixed tripartite states, namely the $\rho_{GW}$, $\rho_{WrG}$, and $\rho_{WrW}$ states, the coherence dynamics depend on both the mixing parameter and the environmental configuration. In the Werner-type states, increasing the weight of the coherent pure-state component reduces the admixture of the maximally mixed state and consequently increases the initial basis-independent coherence. In contrast, the $\rho_{GW}$ mixture exhibits a non-monotonic dependence on the mixing parameter, with the equal mixture displaying the lowest coherence. This behaviour reflects the different nature of the two types of mixtures: the Werner-type states involve admixture with the maximally mixed state, whereas the $\rho_{GW}$ state is a statistical mixture of two distinct coherent pure states. Among the mixed states, the $\rho_{WrW}$ state exhibits particularly strong robustness, especially under common and non-Markovian dephasing environments.\\

\noindent We have also examined a partial-exposure configuration in which two qubits are subjected to local Markovian dephasing while the third qubit is isolated from the environment. The results show that protecting one qubit generally increases the residual basis-independent coherence at the final evolution time compared with the configuration in which all three qubits are exposed to local dephasing. This enhancement is particularly pronounced for the $|WW\rangle$ and $|Star\rangle$ states among the pure states. The same trend persists for the mixed states for all the considered values $p=0.1$, $0.5$, and $0.9$, although the magnitude of the enhancement depends on the structure of the mixed state and the mixing parameter. Thus, even partial isolation from the dephasing environment provides an effective means of improving the preservation of intrinsic multipartite coherence.\\

\noindent Overall, the results demonstrate that environmental memory, collective interactions and partial protection can substantially suppress coherence degradation in tripartite systems. The comparison with the relative entropy of coherence highlights the conceptual distinction between basis-dependent and basis-independent descriptions of coherence. While the relative entropy of coherence can show appreciable degradation and a strong dependence on the chosen reference basis, the basis independent measure considered here reveals large residual coherence and pronounced coherence-freezing behaviour, particularly in non-Markovian environments. The comparison further emphasizes that basis-dependent and basis-independent measures capture different aspects of quantum coherence and can therefore exhibit substantially different dynamical behaviour under the same physical evolution.\\

\noindent These findings indicate that basis-independent coherence provides a useful framework for characterizing intrinsic quantum coherence in multipartite open systems and for assessing its robustness against dephasing noise. The results may be relevant to coherence-preserving quantum protocols in situations where the choice of computational basis is arbitrary or physically irrelevant. The present study is restricted to pure-dephasing environments; extending the analysis to
other physically relevant noise models, including amplitude damping, generalized amplitude damping, and depolarizing channels, constitutes a natural direction for future investigation.

\section*{Acknowledgements}
\noindent S.R. and A.M. acknowledge the support of Techno Main Salt Lake, Kolkata, India. The authors also acknowledge Prof. Md Manirul Ali of Chennai Institute of Technology for fruitful discussions. S.R. acknowledges Prof. Chandrashekar Radhakrishnan for introducing the author to the Basis-Independent coherence measure.

\section*{Data Availability Statement}

\noindent The data supporting the findings of this study are available with the authors.

\section*{Declaration of conflict of interest}
\noindent The authors declare that they have no known competing financial interests or personal relationships that could have appeared to influence the work reported in this paper.\\

\noindent \textbf{Author Contributions:} S.R. developed the idea and prepared the manuscript. A.M. generated the dataset and  plotted the figures using MATLAB. 

\end{document}